# Analyzing Powers and Spin Correlation Coefficients for *p+d* Elastic Scattering at 135 and 200 MeV

B. v.Przewoski a,j), H.O. Meyer a), J.T. Balewski a), W.W. Daehnick c)†, J. Doskow a), W. Haeberli b), R. Ibald a), B. Lorentz e), R.E. Pollock a), P.V. Pancella d), F. Rathmann f), T. Rinckel a), Swapan K. Saha c)h), B. Schwartz b), P. Thörngren-Engblom g), A. Wellinghausen a), T.J. Whitaker a) and T. Wise b) and H. Witała i)

a. Indiana University and Cyclotron Facility , Bloomington, IN 47405
b. University of Wisconsin-Madison, Madison, WI, 53706
c. Univ. of Pittsburgh, Pittsburgh, PA 15260
d. Western Michigan University, Kalamazoo, MI, 49008
e. Forschungszentrum Jülich, Germany
f. Universität Erlangen-Nürnberg, Erlangen, Germany
g. Uppsala University, Uppsala, Sweden
h. Bose Institute, Calcutta, India
i. Jagellonian University, Cracow, Poland
j. E-mail: przewoski@iucf.indiana.edu

16-Sep-2003

*Abstract:* The proton and deuteron analyzing powers and 10 of the possible 12 spin correlation coefficients have been measured for *p+d* elastic scattering at proton bombarding energies of 135 and 200 MeV. The results are compared with Faddeev calculations using two different NN potentials. The qualitative features of the extensive data set on the spin dependence in *p+d* elastic scattering over a wide range of angles presented here are remarkably well explained by two-nucleon force predictions without inclusion of a three-nucleon force. The remaining discrepancies are, in general, not alleviated when theoretical three-nucleon forces are included in the calculations.

## I. Introduction

During the last five years, proton-deuteron scattering has been studied in a number of experiments at intermediate-energy facilities, including RIKEN [1, 2, 3], the KVI [4,5], and IUCF [6,7]. The declared purpose of all of these experiments was the search for evidence of a three-nucleon force.

This considerable experimental activity was stimulated by the availability of parameter-free and computationally exact predictions of scattering observables in the three-nucleon system, derived from a given nucleon-nucleon potential. These "Faddeev" calculations, carried out mainly by the Bochum-Cracow group [8], are now available at intermediate energies, owing to advances in computing power that made the inclusion of a sufficient number of partial waves possible. However, pion production (above ~200 MeV proton energy) is not included in these calculations.

It is commonly argued that discrepancies between data and calculations are a manifestation of physics that is omitted in these calculations, and that the most obvious contender is the three-nucleon force (3NF). Bombarding energies above 100 MeV are of interest because 3NF effects are expected to grow with increasing energy, and because the Coulomb interaction is of minor importance, making it feasible to compare the calculations (which are really for *n+d* scattering) to *p+d* scattering data.

It is also possible to include model representations of the 3NF in the Faddeev calculations. If this were to lead to a systematic improvement of the agreement with the data, one would have uncovered evidence for a 3NF.

Polarization observables contain sums of interfering pairs of amplitudes and are potentially more sensitive than the cross section to contributions from a small effect such as the 3NF. In order to test the present (and any future) models of a 3NF, it is crucial to have measured as many polarization observables as possible. The experiments cited above cover the cross section, the proton analyzing power, the four deuteron analyzing powers, and in one case [2], polarization transfer coefficients. Only a single spin correlation coefficient measurement (beam *and* target polarized) has been reported [7]. In this paper, we report the measurement of 10 of the 12 possible spin correlation coefficients, in addition to the five analyzing powers. The measurement was carried out at 135 and 200 MeV proton bombarding energy, and used a polarized proton beam and a vector- and tensor-polarized deuteron target.

The paper is organized as follows. In Sec. II we define the measured observables and derive the spin-dependent scattering cross section. Sects. III and IV we describe the equipment and the measurement. In Sec. V we explain how the observables were deduced from the data and present the results. In Sec. VI we describe the calculations and present models of the 3NF and compare them to the measurements. This is followed by our conclusions in Sec. VII.

## II. Observables
### A. Coordinate Frames and Definition of Observables
The following discussion is limited to the tools that are needed to analyze the data of this experiment; details of the polarization formalism and its foundation can be found, e.g., in Ohlsen's discussion of spin correlation experiments involving particles with spin ½ and 1 [9]. For the treatment of spin-1 polarization, two different bases are in common use and various normalization conventions can be found in the literature. Here, we are using the Cartesian basis (as opposed to the spherical tensor basis) because it is more intuitive when dealing with spin correlation coefficients. For normalization we follow the Madison

Convention [10]. The production and description of polarized beams is also well explained in ref. [11].

We define as the 'scattering frame' a Cartesian coordinate system (**X,Y,Z**) with the *Z* axis along the momentum of the incident proton, $\vec{p}_{inc}$, the *Y* axis in the direction of $\vec{p}_{inc} \times \vec{p}_{out}$ where $\vec{p}_{out}$ is the momentum of the scattered proton, and the *X* axis completing a right-handed coordinate frame. The differential cross section $\sigma$ for elastic scattering of polarized protons from polarized deuterons, in units of the unpolarized differential cross section $\sigma_0$, is given by Eq. 6.8 of ref. [9] as follows

$$\begin{aligned}
\sigma/\sigma_0 = &\ 1 + Q_Y A_y^p + \tfrac{3}{2} P_Y A_y^d + \\
&+ \tfrac{2}{3} P_{XZ} A_{xz} + \tfrac{1}{3}\left(P_{XX} A_{xx} + P_{YY} A_{yy} + P_{ZZ} A_{zz}\right) + \\
&+ \tfrac{3}{2}\left(P_X Q_X C_{x,x} + P_X Q_Z C_{x,z} + P_Y Q_Y C_{y,y} + P_Z Q_X C_{z,x} + P_Z Q_Z C_{z,z}\right) + \\
&+ \tfrac{1}{3}\left(P_{XX} Q_Y C_{xx,y} + P_{YY} Q_Y C_{yy,y} + P_{ZZ} Q_Y C_{zz,y}\right) + \\
&+ \tfrac{2}{3}\left(P_{XY} Q_X C_{xy,x} + P_{XY} Q_Z C_{xy,z} + P_{XZ} Q_Y C_{xz,y} + P_{YZ} Q_X C_{yz,x} + P_{YZ} Q_Z C_{yz,z}\right) \ .
\end{aligned} \quad (1)$$

Using indices ($I,K = X,Y,Z$) the $Q_I$ are the components of the proton polarization in the scattering frame, the $P_I$ are the components of the deuteron vector polarization and the $P_{IK}$ are the Cartesian moments of the deuteron tensor polarization. The observables, defined by this equation, include the proton analyzing power $A_y^p$, the deuteron vector analyzing power $A_y^d$, the tensor analyzing powers $A_{ik}$, the vector spin correlation coefficients $C_{i,k}$, and the tensor spin correlation coefficients $C_{ik,n}$. These observables are functions of the scattering angle $\theta$. In the derivation of Eq. 1, the constraints of parity conservation have been taken into account. We note that the Cartesian basis is over-complete, and that the following three relations between the terms of Eq. 1 hold

$$P_{XX} + P_{YY} + P_{ZZ} = A_{xx} + A_{yy} + A_{zz} = C_{xx,y} + C_{yy,y} + C_{zz,y} = 0 \ . \quad (2)$$

Defining $A_\Delta \equiv A_{xx} - A_{yy}$ and $C_{\Delta,y} \equiv C_{xx,y} - C_{yy,y}$, we use the relations of Eq. 2 to eliminate $A_{xx}+A_{yy}$ and $C_{xx,y}+C_{yy,y}$. There are then 17 spin observables in all, namely the proton and deuteron vector analyzing power, three tensor analyzing powers, five vector correlation coefficients and seven tensor correlation coefficients.

The task of extracting spin observables from the data requires the measurement of the azimuthal dependence of the cross section, calling for a cylindrically symmetric detector. In such a detector, the azimuth $\varphi$ of the scattering plane around the beam axis is a measured quantity that varies from event to event. To describe this, we define a second Cartesian frame (**x,y,z**) that is *fixed* in space with the *x* axis pointing to the left, the *y* axis upwards and the *z* axis in the beam direction. The azimuth $\varphi$ of the outgoing proton (i.e., the orientation of the scattering plane) is measured clockwise from the positive *x*-axis, looking in the beam direction. Thus, the scattering frame is obtained by rotating the fixed frame by $\varphi$ around the *z* (or, *Z*) axis.

The polarization of the (spin-½) proton beam is specified in the fixed frame by a three-component vector with magnitude $Q$ and direction $\hat{Q} = (\beta_Q, \Phi_Q)$, where $\beta_Q$ is the polar angle (with respect to the *z* axis) and $\Phi_Q$ is the azimuth. The polarization components in the scattering frame are then given by

$$Q_X = Q \sin \beta_Q \cos(\Phi_Q - \varphi) ,$$
$$Q_Y = Q \sin \beta_Q \sin(\Phi_Q - \varphi) , \quad (3)$$
$$Q_Z = Q \cos \beta_Q .$$

The description of the polarization of the deuteron target is more complicated. For an ensemble of spin-1 particles prepared by an atomic beam source there exists an axis of rotational symmetry $\hat{S}$, called "spin alignment axis". Let us denote by $m_+$, $m_0$ and $m_-$ the fractional populations of the three magnetic substates with projection +1, 0, and –1 with respect to a quantization axis in the direction of $\hat{S}$. The vector polarization of the ensemble is then given by $P_\zeta = m_+ - m_-$ and the tensor polarization by $P_{\zeta\zeta} = 1 - 3m_0$. In order to characterize the polarization of the deuteron target, the orientation of the spin alignment axis, $\hat{S} = (\beta_P, \Phi_P)$, must be known, in addition to the values of $P_\zeta$, $P_{\zeta\zeta}$. The spin alignment axis is associated with the expectation value of the magnetic moment (either parallel or anti-parallel) and thus can be controlled by the guide field at the target as explained in Sec. III.3. The components of the *vector* polarization are analogous to the proton case,

$$P_X = P_\zeta \sin \beta_P \cos(\Phi_P - \varphi) ,$$
$$P_Y = P_\zeta \sin \beta_P \sin(\Phi_P - \varphi) , \quad (4)$$
$$P_Z = P_\zeta \cos \beta_P ,$$

while the tensor moments are given by [9,11]

$$P_{XY} = \tfrac{3}{4} P_{\zeta\zeta} \sin^2 \beta_P \sin 2(\Phi_P - \varphi) ,$$
$$P_{YZ} = \tfrac{3}{2} P_{\zeta\zeta} \sin \beta_P \cos \beta_P \sin(\Phi_P - \varphi) ,$$
$$P_{XZ} = \tfrac{3}{2} P_{\zeta\zeta} \sin \beta_P \cos \beta_P \cos(\Phi_P - \varphi) , \quad (5)$$
$$P_\Delta \equiv P_{XX} - P_{YY} = \tfrac{3}{2} P_{\zeta\zeta} \sin^2 \beta_P \cos 2(\Phi_P - \varphi) ,$$
$$P_{ZZ} = \tfrac{1}{2} P_{\zeta\zeta} (3\cos^2 \beta_P - 1) .$$

**B. Polarized Cross Section**

We start from Eq.1, eliminate the dependent variables using Eq. 2 and insert Eqs. 3-5. This leads to an equation for $\sigma/\sigma_0$ that contains the values for beam and target polarization, $Q$, $P_\zeta$, $P_{\zeta\zeta}$, the orientations of beam polarization vector $\hat{Q}$ and of the target spin alignment axis $\hat{S}$, the observables, and the azimuth $\varphi$ of the scattering plane. At this stage it is practical to evaluate the cross section for those specific orientations $\hat{Q}(\beta_Q, \Phi_Q)$ and $\hat{S}(\beta_P, \Phi_P)$ that are actually used in this experiment.

We used different scenarios for beam and target polarization. In scenario V90 (see Sec. IV.1.2) the beam polarization was *vertical* (along the $y$ axis), thus $\beta_Q = \pi/2$, and $\Phi_Q = \pi/2$. For a *sideways* deuteron spin alignment axis $\hat{S}$, we have $\beta_P = \pi/2$, and $\Phi_P = 0$. Eq.1 then reduces to

$$\sigma/\sigma_0 = 1 + QA_y^p \cos\varphi - \tfrac{3}{2} P_\zeta A_y^d \sin\varphi - \tfrac{1}{4} P_{\zeta\zeta}\left[A_{zz} - A_\Delta \cos 2\varphi\right]$$
$$+ \tfrac{3}{4} P_\zeta Q\{C_{x,x} - C_{y,y}\}\sin 2\varphi \qquad (6)$$
$$- \tfrac{1}{4} P_{\zeta\zeta} Q\left[\{C_{zz,y} + (C_{xy,x} - \tfrac{1}{2}C_{\Delta,y})\}\cos\varphi - \{C_{xy,x} + \tfrac{1}{2}C_{\Delta,y}\}\cos 3\varphi\right].$$

In deriving this equation, when products and powers of trigonometric functions of $\varphi$ occur, they are transformed to expressions containing only members of the orthogonal set $\cos(k_c\varphi)$ ($k_c = 0,1,2\ldots$) and $\sin(k_s\varphi)$ ($k_s = 1,2\ldots$). On the other hand, for a *vertical* deuteron spin alignment axis we have $\beta_P = \pi/2$, and $\Phi_P = \pi/2$), and we obtain

$$\sigma/\sigma_0 = 1 + QA_y^p \cos\varphi + \tfrac{3}{2} P_\zeta A_y^d \cos\varphi - \tfrac{1}{4} P_{\zeta\zeta}\left[A_{zz} + A_\Delta \cos 2\varphi\right]$$
$$+ \tfrac{3}{4} P_\zeta Q\left[\{C_{x,x} + C_{y,y}\} - \{C_{x,x} - C_{y,y}\}\cos 2\varphi\right] \qquad (7)$$
$$- \tfrac{1}{4} P_{\zeta\zeta} Q\left[\{C_{zz,y} - (C_{xy,x} - \tfrac{1}{2}C_{\Delta,y})\}\cos\varphi + \{C_{xy,x} + \tfrac{1}{2}C_{\Delta,y}\}\cos 3\varphi\right],$$

and choosing the deuteron spin alignment axis *along* the beam direction ($\beta_P = 0$), leads to

$$\sigma/\sigma_0 = 1 + QA_y^p \cos\varphi + \tfrac{1}{2} P_{\zeta\zeta} A_{zz} + \tfrac{3}{2} P_\zeta Q C_{z,x} \sin\varphi + \tfrac{1}{2} P_{\zeta\zeta} Q C_{zz,y} \cos\varphi . \qquad (8)$$

During the course of the experiment, the values of $Q$, $P_\zeta$ and $P_{\zeta\zeta}$ can be made positive, negative or zero. This is used to separate terms with vector and tensor polarization, and terms that contain only the beam or the target polarization (analyzing powers), or both (spin correlation coefficients). The remaining decomposition makes use of the known azimuthal dependence of the cross section. It should be pointed out that the actual results of the experiment are the factors associated with the trigonometric functions in Eqs. 6-8. In some cases these are linear combinations of spin observables. Inspecting Eqs. 6-8 one sees that these combinations can be combined to extract the following observables:
$A_y^p$, $A_y^d$, $A_\Delta$, $A_{zz}$, $C_{x,x}$, $C_{y,y}$, $C_{z,x}$, $C_{zz,y}$, $C_{xy,x}$, and $C_{\Delta,y}$.

Other choices of the polarization directions (see Sec. IV.1) are treated in an analogous fashion. The resulting spin-dependent cross sections are given in the appendix.

### III. Experimental equipment
**A. Overview**
This experiment makes use of a stored, polarized proton beam in the Indiana Cooler. The experiment is located in the A-region of the Cooler where the dispersion almost vanishes and the horizontal and vertical betatron functions are small [12], favoring the use of a narrow target cell. The target setup (Fig.1, a-d) consists of an atomic beam source [13,14] that injects polarized deuterium atoms into a storage cell. The proton and the deuteron from elastic scattering are detected in coincidence by a detector system consisting of scintillators, wire chambers (j-m) and recoil detector array (e) surrounding the target cell.

**B. Polarized Proton Beam**
*1. Beam Properties*
Protons are produced by a polarized ion source, accumulated in the injector synchrotron and then injected into the Cooler. About ten transfers at 1 Hz result in a typical stored current of about 500μA. The experiment was carried out at 135 and 200 MeV (the actual beam energies are known to +/-0.1 MeV and have been measured from the orbit frequency and ring circumference to be 135.0 and 203.3 MeV). The beam polarization is typically 0.75; its sign is reversed for every fill of the Cooler. Prior to each fill, the ring is completely emptied by resetting the main magnets. The betatron tunes of the Cooler are adjusted to avoid any depolarizing resonances; the polarization lifetime is then much longer than the beam lifetime.

*2. Longitudinal Beam Polarization*
In the absence of non-vertical fields, the stable spin direction in a circular accelerator is vertical. In order to obtain longitudinal beam polarization at the target, two "spin rotators" (longitudinal magnetic fields) are used [15]. One rotator is introduced by operating all solenoids in the cooling region with the *same* sign. These include the main solenoid that confines the electron beam and two solenoids, immediately upstream and downstream, which are normally used to compensate for the cooling solenoid field. Between the target and the cooling region, the beam is bent by 120°. The other rotator consists of a superconducting solenoid halfway between the target and the cooling region (for details, refer to Ref. [15]). Data with longitudinal beam polarization were taken only at 135 MeV. At this energy, a longitudinal field integral of 0.56 T m for both rotators results in nearly longitudinal polarization with a small (about 0.08) vertical component.

Of the injected beam polarization, only the component that is parallel to the stable spin direction at the injection point is preserved. When the spin rotators are used, the stable spin direction at injection is tilted by about 45º towards the beam direction, i.e., no longer vertical. Thus, an additional solenoid was used in the transfer beam line between injector synchrotron and the Cooler to match the two directions.

**C. Polarized Deuteron Target**
*1. Overview*
The internal, polarized deuteron target is generated by injecting polarized atoms from an atomic beam source (ABS) into a storage cell. The target is placed in a weak guide field generated by a set of Helmholtz-like coils (Fig.1, g, i). A set of similar coils with opposite field (h) practically eliminates a correlated position shift of the stored beam.

In the ABS, atoms from an 18 MHz dissociator (a) emerge through an aluminum nozzle that is kept at liquid nitrogen temperature. The atoms then pass through two stages, each consisting of a set of sextupole magnets (b) followed by a medium field transition unit (c). In the sextupole magnets the atoms are separated according to their electron polarization. In the first medium-field transition unit (MF1), transitions between hyperfine states are induced. After passing through the second set of sextupole magnets, which rejects one of the three hyperfine states present in the beam, another transition between hyperfine states may be induced in the second medium field transition unit (MF2).

For previous operation with *hydrogen*, the ABS had been equipped with a single, fixed-gradient medium-field transition unit located after the first set of sextupole magnets. Operation of the ABS in this configuration is extensively described elsewhere [14]. Here, we concentrate on the description of two new medium-field transition units (c) that were added for operation of the source with deuterium and were used for the first time by this experiment.

*2. Medium-Field Transitions*
A medium-field transition operates in magnetic fields of $0.1B_c$ to $0.2B_c$, where $B_c$ is the hyperfine interaction field of 50.7 mT for hydrogen and 11.7 mT for deuterium. In addition to a uniform (offset) field, a field gradient along the beam direction is required to satisfy the condition of adiabatic passage.

Multiple transitions can be made by adjusting the offset field so that the beam passes in sequence through field regions where the populations of different pairs of hyperfine states are interchanged at a given, fixed RF frequency [16].

In order to enable remote change between different operating modes of the target, two new transition units with variable gradient and variable offset field were installed. The linearity of the gradient field over the transition region as well as the homogeneity of the offset field were measured prior to installation of the units in the ABS.

For deuterium the gradient field is set to +0.2 mT/cm. The RF coil of each MF unit consists of a 70 mm long, 12-turn solenoid with 34 mm diameter, made from 1.6 mm diameter wire. For deuterium, the coils are operated at 60.5 MHz, and for hydrogen at 30 MHz. The transition units are water-cooled. The currents in the offset and gradient coils are remotely controlled. This makes it possible to quickly change between vector, positive tensor, and negative tensor polarization, while data are being acquired. Hall probes are used to monitor the field in the transition units.

## *3. Operation of the Atomic Beam Source*

After the first set of sextupoles the atomic beam consists of states 1+2+3, where the states are labeled in order of decreasing energy in a non-zero magnetic field [17]. Up to three transitions are made sequentially in MF1. The gradient field is kept constant while the offset field is changed for different spin states. For a small offset field no transition is made in MF1. When the offset field is increased, the atoms undergo a 3→4 transition.

When the field is further increased the atoms pass through the 3→4 transition followed by the 2→3 transition. If the offset field is increased even further, the atoms undergo the 3→4, 2→3 and 1→2 transitions sequentially. The second set of sextupoles eliminates state 4, so that one is left with states 1+2+3, 1+2, 1+3 or 2+3 depending on whether none, one, two, or three transitions are made in MF1. The corresponding maximum nuclear polarizations of the atomic beam, before entering MF2, are $(P_\varsigma, P_{\varsigma\varsigma}) = (+1/3, -1/3)$, $(P_\varsigma, P_{\varsigma\varsigma}) = (+2/3, 0)$, $(P_\varsigma, P_{\varsigma\varsigma}) = (+1/3, 0)$ and $(P_\varsigma, P_{\varsigma\varsigma}) = (0, -1)$. MF2 is only needed to produce positive tensor polarization. Then, its parameters are set such that atoms in states 1 and 3 with polarizations $(P_\varsigma, P_{\varsigma\varsigma}) = (+1/3, 0)$ undergo the 3→4 transition. Consequently, after passing through MF2 the atomic beam contains states 1 and 4 with polarization $(P_\varsigma, P_{\varsigma\varsigma}) = (0, +1)$.

## *4. Target Cell*

The target cell (Fig. 1, d) is a 27 cm long tube of 12 mm diameter made from 0.05 mm thick aluminum, through which the stored beam travels, very similar to a design used earlier [18]. The cell is coated with Teflon in order to minimize depolarization by wall collisions [19]. The atomic beam from the ABS enters through a feed tube attached to the side of cell. The length of the cell between the feed tube and the downstream end is 12.5 cm; the upstream part is 14.5 cm long. The cell is supported at the intake of the feed tube (away from the beam), minimizing obstructions in the path of the scattered particles. Routinely, the target thickness is about $10^{13}$ atoms/cm$^2$.

The target cell is centered within an array of Helmholtz-like coils that provide horizontal, vertical and longitudinal guide fields of about 0.3 mT for alignment of the target polarization [13,20]. Certain polarization observables require that the angle of the spin alignment axis is at $\beta=45^o$ with respect to the beam. This is achieved by simultaneously exciting either the vertical and longitudinal coils, or the horizontal and longitudinal coils.

*5. Spin Exchange*

The measured values for both, vector and tensor, target polarizations were about 0.45. This means that the tensor polarization is less than half and the vector polarization only about 70% of the theoretical maxima (1.0 and 2/3, respectively). Some decrease from the maximum values can be expected from wall depolarization, incomplete rejection of unwanted states by the sextupoles and an inefficiency of the transition units.

However, in a dedicated measurement [21] we also found that the tensor polarization decreases with increasing target thickness, while, at the same time, the vector polarization shows no such dependence. This behavior is consistent with the loss of polarization due to spin exchange between the deuterium atoms in the cell. A model calculation of the effect of spin exchange [22] explains the observed tensor polarization in a weak magnetic field as a function of target density.

**D. Unpolarized Target**

The procedure to calibrate the beam polarization (see Sec.V.2), calls for an unpolarized, mixed hydrogen and deuterium target. To this aim, an $H_2$-$D_2$ gas mixture is prepared by filling an empty cylinder with approximately equal parts of hydrogen and deuterium (one does not have to know the exact mixing ratio for the calibration). The gas mixture is admitted to the cell through a thin (1 mm diameter) Teflon hose, connected to a nipple at the center of the cell at a rate comparable to the flux of atoms from the ABS.

**E. Detector System**

*1. Overview*

The outgoing proton and deuteron from $p+d$ elastic scattering are detected in coincidence. The detector setup is shown in Fig.1. Most of the components of the detector have been used previously and are described in detail in ref. [18].

*2. Forward Detector*

The forward going particle is detected in a stack consisting of a $\Delta E$ ("F") detector (Fig.1, j), two wire chambers (k,l) with two wire planes each, and a stopping ("K") detector (m).

The F-detector is made from organic scintillator material, segmented into an upper and a lower half. Its initial thickness of 1.5 mm has been increased to 6.4 mm during the course of the experiment. The thicker detector improves the mass resolution for particle identification. The two wire chambers are positioned 22.4 cm and 30.2 cm from the target center and have a wire spacing of 3.2 mm and 6.4 mm, respectively. The K-detector is made from 15.2 cm thick scintillator, segmented into four quadrants. The forward detector system covers the laboratory polar angles between $10^o$ and $45^o$.

*3. Recoil Detector*

The recoil particle is detected in a so-called silicon barrel (Fig.1, e) that consists of an array of eighteen silicon strip detectors [23] surrounding the target cell. Fig. 2 shows the silicon barrel with the target cell in its center. The strips are oriented in such a way that they measure the azimuth of the recoil with a resolution of 2°. The silicon detectors yield an energy measurement from the back plane and a logic signal for each strip on the front plane. Energy and time are read out for each individual detector, but the strips at the same azimuth for a group of three detectors along the beam are electrically connected to reduce the number of electronics channels. The detector with the hit is identified from the energy signal. The silicon detectors are calibrated periodically using an array of six low-level (nCi) $^{241}$Am sources, mounted at the upstream end of the silicon barrel. Each source is positioned to illuminate one of the six sides of the barrel.

The active area of each detector is 4x6 cm$^2$. The downstream ring consists of six 500 μm thick detectors while all other detectors are 1000 μm thick. The detectors are operated at full depletion and cooled to about 0° C.

It has been found that exposure to atomic deuterium or hydrogen has a detrimental effect on silicon detectors. Even a short exposure (30 min) to ambient atomic deuterium causes an increase in leakage current that renders the detectors useless for data acquisition. To prevent atomic deuterium that is leaking from pinholes in the cell from reaching the detectors, the target cell is placed in a bag made from thin Kapton. In addition, copper recombination baffles are placed around the feed tube and at the ends of the barrel. On a copper surface, atoms recombine into harmless molecular deuterium. In this way, the effect of atomic deuterium can be reduced to manageable proportions. Fortunately, the effect of atomic deuterium on the detectors is reversible. Thus, while no longer exposed to atomic deuterium, i.e., between runs, the detectors recovered.

## IV. Measurement
### A. Cycle Time Scenarios
#### 1. Definitions, Parameters Varied
A "cycle" is the time between fills of the Cooler with beam. Proton beam of opposite polarization is injected for alternating cycles. After the fill, the experiment is enabled for data taking. The operating parameters (guide fields and transition units) of the target are varied *during* the cycle in order to acquire data with different target polarizations, but with the *same* stored beam. This is invaluable in minimizing systematic errors.

The guide field that determines the spin alignment axis of the deuteron target is changed in 2 s intervals. The normal sequence includes the six directions left (+*x*), right (–*x*), down (–*y*), up (+*y*), along (+*z*) and opposite (–*z*) to the beam axis. We call this a "sub-cycle". Note that a sign change of the guide field affects the vector, but not the tensor polarization.

Vector or tensor polarization of the target is selected by enabling different sets of transitions (Sec. III. 3.3) by remotely changing the offset field in the transition units, while keeping the gradient field constant. To overcome the effects of hysteresis, the transition units are de-gaussed before any change. This is accomplished by applying a 2 Hz alternating current with exponentially decreasing amplitude to all transition-unit coils. De-gaussing takes about 5 s (see Fig. 3).

In the following we describe the three cycle-time scenarios used in this experiment.

## 2. Scenario V90

In scenario V90 the beam polarization is vertical. The target guide field is along the *x* or *y* axis ($\beta_P = 90°$), or the *z* axis ($\beta_P = 0°$). Within each cycle, the state of the atomic beam source is set to positive tensor polarization for two normal sub-cycles, to vector polarization for two sub-cycles, and finally to negative tensor polarization for three sub-cycles. Negative tensor is measured longer to approximately compensate for the loss in intensity due to the finite beam lifetime. Note, that both signs of vector polarization are available because the guide field changes sign during the sub-cycle.

Fig. 3 shows three selected quantities measured during a V90 cycle. The top panel illustrates the beam current in the ring. The current in the offset field coil in transition unit MF1 is shown in the middle panel. One can see the three current plateaus (positive tensor, vector, negative tensor), each preceded by the de-gaussing of the coil. The event rate during data taking is depicted in the bottom panel. During de-gaussing, no transitions are made, admitting an additional sub-state to the target cell; thus, the target thickness and therefore the event rate increase during de-gaussing.

A total of 5662 (7737) V90 cycles were acquired at 135 (200) MeV.

## 3. Scenario V45

The purpose of scenario V45 is to measure observables that require a deuteron spin alignment axis that is *not* along the axes of the coordinate frame. To this aim, a sub-cycle is used for the guide fields in which *two* sets of coils are energized simultaneously, the corresponding magnetic field directions adding vectorially. This special sub-cycle consists of the eight states (+x,+z), (+x,–z), (–x,+z), (–x,–z), (+y,+z), (+y,–z), (–y,+z), and (–y,–z). This corresponds to orientations of the deuteron spin alignment axis at angles $\beta_P = 45°$ or $135°$, either in the horizontal or the vertical plane. Again, these states are changed every 2 s. The atomic beam source is set in turn to positive polarization for two special sub-cycles and negative tensor polarization for three sub-cycles. Vector polarization is not used in scenario V45. The beam polarization is also vertical.

A total of 2317 (1873) V45 cycles were acquired at 135 (200) MeV.

## 4. Scenario L90

The purpose of scenario L90 is to measure some observables that require longitudinal beam polarization (see Tab. 1). During the whole cycle the target is vector-polarized, and a normal sub-cycle is used as in scenario V90. Scenario L90 is used only at 135 MeV (a series of power outages is responsible for the lack of data at the higher energy).

A total of 1905 L90 cycles were acquired.

**B. Event Sorting**

The goal of event sorting is to select $p+d$ elastic scattering events using the signals generated by the detectors. The condition that triggers the readout of the entire detector is a coincidence between the upper half of the K-detector and the lower half of the silicon barrel, or *vice versa*.

For each event, the angles of the forward prong ($10° < \theta_{lab} < 45°$, $0° < \varphi < 360°$) are determined from the wire chambers. Normally there is one hit in each of the four wire chamber planes, however, events with one plane missing or with two hits in one or two planes can be reconstructed and are also used. The angular resolutions estimated from the wire spacing are $\delta\theta_{lab} = 2.2°$ and $\delta\varphi = 2.6°$.

The gains of all scintillator tubes are corrected in software for shifts due to different guide fields in order to eliminate spin dependence of the detector performance. Also corrected are the position dependence of the light collection efficiency and the time response of the F- and the K-detectors. For more details, see Ref. [18].

The forward particle can be either a proton or a deuteron. At 135 MeV incident energy both particles stop in the K-detector, while at 200 MeV only the deuteron is stopped. Particle identification makes use of the correlation between the deposited energies in the F- and the K- detector (Fig. 4, A), as well as the correlation between F-K time-of-flight and the deposited energy in the K- detector (B). To further discriminate against background from breakup events, additional gates are placed on the correlation between the scattering angle and energy deposited in the K-detector (C), consistent with elastic scattering kinematics, and the correlation between energy deposited in the silicon detector and the scattering angle of the forward prong (D).

The silicon detectors measure the azimuth of the recoil with a resolution of $2°$. Events where a single strip or a pair of adjacent strips fires are accepted in the analysis. This determines the azimuth of the recoil, and thus the difference $\Delta\varphi$ between the two prongs. Elastic scattering events, being coplanar, are required to have $\Delta\varphi$ between $175°$ and $185°$.

The center-of-mass-angle $\theta$, calculated from the forward lab angle, is sorted into $4°$ wide bins, and the azimuth $\varphi$ into $12°$ bins. After applying all software conditions, two-dimensional ($\theta$ versus $\varphi$) arrays of yields are generated for each spin state, including all combinations of two signs of the beam polarization, target vector, positive tensor or negative tensor, and six (scenarios V90, L90) or eight (scenario V45) guide field directions. A software gate on the cycle number versus cycle time is used to eliminate incomplete sub-cycles in order to reduce spin-dependent luminosity corrections.

**C. Background**

One expects that unwanted background events arise mainly from $p+d$ breakup. In order to assess the effect of background on the spin observables, we study the distribution of the difference $\Delta\varphi$ between the azimuths of the forward and recoil particle. Fig. 5 shows this distribution after all other cuts have been applied. One sees that the coplanar peak at 180° from elastic scattering is superimposed on a wider distribution, which we associate with background. For good events, $\Delta\varphi$ is required to fall between 175° and 185°. In order to generate a background-enriched event sample, we instead select the wings with 50° < $\Delta\varphi$ < 150° and 210° < $\Delta\varphi$ < 310°, and repeat the process of event sorting with the same conditions as for good events, except for the coplanarity requirement. From the resulting yields we then deduce background-enriched observables.

The amount of background (5-10%) under the $\Delta\varphi$ peak is determined from a smooth approximation of the wings (solid line in Fig. 5). Assuming that the observables associated with the background under the peak are the same as for the background in the wings, it is straightforward to calculate a background correction for the good data. This is done for all $\theta$ bins separately. We find that these corrections for all observables at all angles are smaller than the statistical errors in all cases, reflecting the fact that the observables from events in the peak or in the wings are very similar. Thus, it seems that the event conditions discriminate rather well against $p+d$ breakup, and that the events in the $\Delta\varphi$ wings are not background at all, but real events in the tail of the angular resolution.

We conclude that corrections due to background are negligible. This conclusion is supported by an analysis of the cross section, discussed in Sec. V.4.

**D. Corrections**
*1. Geometric Corrections*
The wire chambers define the coordinate frame of the experiment. Their positions have been surveyed optically prior to the experiment. The beam position, which may vary for different setups of the Cooler ring, can be extracted from the distribution of the event vertex positions. The original wire chamber coordinates are then offset such that the beam coincides with the *z* axis. The magnitude of the offset was always less than 1.5 mm.

The scattering angle is determined from the intercept of the forward track with the two wire chambers. The distance between the chambers affects the absolute value of this angle. A small correction to the wire chamber positions is applied such that the zero transitions of the vector analyzing power at 135 MeV [3] at forward and backward angles are reproduced.

With the wire chamber offsets known, the positions of the silicon detectors are determined. For each silicon detector three parameters are adjusted, namely the *x* and *y* coordinates of the center of strip 1 and an angle of rotation about the strip direction. These parameters remained the same throughout the experiment, unless a detector was replaced. In addition, overall *x* and *y* offsets of the entire barrel are determined to account for shifts of the beam position (usually accompanying an energy change) by requiring that the difference in azimuth, $\Delta\varphi$, between the forward and the backward prongs peaks at 180°.

## 2. Spin-Dependent Deadtime

In the case of longitudinal beam polarization the trigger rate may depend on the alignment of beam and target spin, which may translate into a spin-dependent deadtime. When the deadtime of the acquisition system, determined from the ratio of triggers issued and processed, is sorted according to spin states, a small dependence of the deadtime on the relative alignment of beam and target spin is found. Correcting the measured yields accordingly results in a small offset (0.026) to $C_{z,z}$, which is measured only at 135 MeV. All other observables are unaffected by deadtime.

# V. Data analysis
## A. Extraction of Observables from Spin-sorted Yields
### 1. Spin-dependent Yields

Throughout this experiment, the proton beam polarization is either vertical or longitudinal and its sign is alternated every cycle. In addition, the target polarization (vector or tensor, guide field direction) is varied, during the cycle, according to three different scenarios (Sec. IV.1.1). For each combination of the beam and target parameters, the event sorting (Sec. IV.2) results in yields $Y$ (or, number of events), stored in an array as a function of $\theta$ (4° bins) and $\varphi$ (12° bins).

### 2. Extracting Observables

We make the following assumptions:

(i) The magnitude of the target polarization does not depend on the direction of the guide field. This has been verified to a high degree of precision ($\pm 0.005$) in previous measurements with this apparatus [13]. For guide fields of opposite sign, the vector polarization has opposite sign, but the tensor polarization stays the same.

(ii) The integrated luminosity in two target states of opposite sign of the target field is the same. A possible difference that arises from the decrease of the beam intensity by about 0.1% per second is negligible.

(iii) The ratio of the luminosities acquired with positive and negative tensor target polarization is the same for both signs of the beam polarization.

(iv) When the target is vector-polarized, the tensor polarization vanishes (verified during commissioning of the transition units). The converse, admixture of vector polarization to a tensor target, is of no concern since in the analysis of tensor terms, vector terms cancel because of the changing sign of the guide field.

We do *not* assume that the magnitudes of opposite-sign beam polarization and of opposite-sign target *tensor* polarization are the same, or that data with *equal* integrated luminosity have been acquired with opposite sign of beam and target *tensor* polarization, since in the present experiment this is not strictly the case. However, we start with the concept of an *ideal* experiment, where these conditions would also be fulfilled, and introduce departures from an ideal experiment as corrections.

### 3. Asymmetries

We select four yields, $Y_{++}$, $Y_{+-}$, $Y_{-+}$, $Y_{--}$, where the first sign refers to the sign of the beam polarization, and the second to the sign of the target polarization. This can be done either for the vector or the tensor target. From the four yields we form the following three ratios, henceforth called *asymmetries*.

$$R_Q = \frac{(Y_{++} + Y_{+-}) - (Y_{-+} + Y_{--})}{Y_{++} + Y_{+-} + Y_{-+} + Y_{--}} \tag{9}$$

$$R_P = \frac{(Y_{++} + Y_{-+}) - (Y_{+-} + Y_{--})}{Y_{++} + Y_{-+} + Y_{+-} + Y_{--}} \tag{10}$$

$$R_{QP} = \frac{(Y_{++} + Y_{--}) - (Y_{-+} + Y_{+-})}{Y_{++} + Y_{--} + Y_{-+} + Y_{+-}} \tag{11}$$

For an ideal experiment, $R_Q$ only depends on the beam polarization, $R_P$ only on the target polarization, while the correlation asymmetry $R_{QP}$ depends on both. In these ratios, the detector efficiency cancels, and thus azimuthal variations in efficiency disappear. Like the yields, the asymmetries $R$ are functions of $\theta$ and $\varphi$.

In scenario V90, there are the three guide-field directions, $B_x$, $B_y$ and $B_z$ (sideways, vertical and longitudinal), and data are taken with a vector or a tensor target. Thus, there are 18 asymmetries. An example of the $\varphi$-dependences of these asymmetries is shown in Fig. 6. These $\varphi$ distributions form the basis for the extraction of the observables.

It is straightforward to express the asymmetries in terms of the observables by inserting the expressions for the polarized cross section into Eqs. 9-11. The beam asymmetry is independent of the target state and given by

$$R_Q = Q\, A_y^p \cos\varphi \quad . \tag{12}$$

The target asymmetries $R_P$ and the correlation asymmetries $R_{PQ}$ depend on the direction of the guide field ($x$, $y$, and $z$, indicated by a superscript) and on whether the target is vector ($P_\zeta$) or tensor ($P_{\zeta\zeta}$) polarized. The target asymmetries are then given by

$$R^x_{P_\zeta} = -\tfrac{3}{2}\, P_\zeta A_y^d \sin\varphi \;, \tag{13}$$

$$R^y_{P_\zeta} = \tfrac{3}{2}\, P_\zeta A_y^d \cos\varphi \;, \tag{14}$$

$$R^x_{P_{\zeta\zeta}} = -\tfrac{1}{4} P_{\zeta\zeta} [A_{zz} - A_\Delta \cos 2\varphi] \;, \tag{15}$$

$$R^y_{P_{\varsigma\varsigma}} = -\tfrac{1}{4} P_{\varsigma\varsigma} [A_{zz} + A_\Delta \cos 2\varphi] , \qquad (16)$$

$$R^z_{P_{\varsigma\varsigma}} = \tfrac{1}{2} P_{\varsigma\varsigma} A_{zz} , \qquad (17)$$

and the correlation asymmetries by

$$R^x_{P_\varsigma Q} = \tfrac{3}{4} P_\varsigma Q (C_{x,x} - C_{y,y}) \sin 2\varphi , \qquad (18)$$

$$R^y_{P_\varsigma Q} = \tfrac{3}{4} P_\varsigma Q [(C_{x,x} + C_{y,y}) - (C_{x,x} - C_{y,y})] \cos 2\varphi , \qquad (19)$$

$$R^z_{P_\varsigma Q} = \tfrac{3}{2} P_\varsigma Q C_{z,x} \sin \varphi , \qquad (20)$$

$$R^x_{P_{\varsigma\varsigma} Q} = -\tfrac{1}{4} P_{\varsigma\varsigma} Q [(C_{zz,y} + (C_{xy,x} - \tfrac{1}{2} C_{\Delta,y})) \cos \varphi - (C_{xy,x} + \tfrac{1}{2} C_{\Delta,y}) \cos 3\varphi] , \qquad (21)$$

$$R^y_{P_{\varsigma\varsigma} Q} = -\tfrac{1}{4} P_{\varsigma\varsigma} Q [(C_{zz,y} - (C_{xy,x} - \tfrac{1}{2} C_{\Delta,y})) \cos \varphi + (C_{xy,x} + \tfrac{1}{2} C_{\Delta,y}) \cos 3\varphi] , \qquad (22)$$

$$R^z_{P_{\varsigma\varsigma} Q} = \tfrac{1}{2} P_{\varsigma\varsigma} Q C_{zz,y} \cos \varphi . \qquad (23)$$

Comparison of these expressions with Fig. 6 shows that the expected $\varphi$-dependences are borne out nicely by the data. The values for the observables times the respective polarizations (henceforth called "asymmetry terms") are then extracted from the yields by fitting simple trigonometric functions (Eqs. 12–23) to the $\varphi$-dependence (solid curves in Fig. 6). This procedure is carried out for each polar angle bin. The primary measured quantities are thus these asymmetry terms. Note, that in some cases asymmetry terms are linear combination of observables.

The statistical errors are derived from the errors $\delta Y^2 = Y$ of the yields in Eqs. 9 – 11 by standard error propagation, neglecting covariance terms. This is justified since the yields are the result of separate experiments and taken at interleafed, but different times. The same is true for the $R$'s in Eqs. 13 – 23, which are obtained with different states of the polarized source or the target field. The asymmetry terms follow from a fit to $\varphi$ distributions where each bin corresponds to a different part of the detector.

## 4. Departure from an Ideal Experiment

*Alignment of the polarization directions.* The coordinate axes of the experiment are defined by the wire chambers, while the beam polarization direction is given by the spin closed orbit and the target polarization direction by the guide fields. Discrepancies between these three frames are taken into account by a shift $\delta\varphi$ of the azimuth scale. This shift is easily determined by comparing the data with the predicted $\varphi$ dependence. For the target orientation we find $\delta\varphi = -3°$, while the beam orientation is shifted by $\delta\varphi = -6°$ at 135 MeV and by $\delta\varphi = -3.5°$ at 200 MeV. The error in determining the φ offsets is ± 0.5°. A small correction term is introduced in the analysis that takes into account that the orientations of target and beam are slightly different.

*Differences in polarization and luminosity of states of opposite polarization.* Beam polarization of opposite sign is produced with different transition units in the ion source and it is not guaranteed that the two polarizations have the same magnitude. The imbalance $q$ (the difference divided by the sum) varies from run to run and is typically 10%. Similarly, target tensor polarization of opposite sign uses different transitions in the ABS. The imbalance $p$ in this case is 1-2%. The relative luminosities with beam of opposite sign may also differ, but when averaged over many cycles, the corresponding imbalance $\mu$ is typically small (1%). The largest departure from an ideal experiment arises from the difference in luminosity with the tensor target states of opposite sign, occurring at different times in the cycle. There is systematic imbalance $\eta$ of about 18%, consistent with the beam lifetime. All four imperfection parameters, $q$, $p$, $\mu$ and $\eta$ can be deduced from the data. Once they are known, the yield equations are worked out including new terms that depend on these parameters. Ignoring higher-order terms, this leads to a system of linear equations between the non-ideal (measured) asymmetries and their corresponding ideal values. The latter are deduced and used in the analysis described in the preceding section.

## *5. Results from Different Scenarios*

So far, we have described the method of analysis for scenario V90. The same principle is used to deduce observables from runs under scenarios V45 and L90 (the corresponding cross sections are given in the appendix). Scenario V45 (Sec. IV.1.3) uses a deuteron spin alignment axis bisecting the *x*- and *z*-axes, or the *y*- and *z*-axes, and scenario L90 (Sec.IV.1.4) employs longitudinal beam polarization. Because longitudinal polarization is accompanied by a small vertical component, this measurement is also sensitive to some of the terms measured in scenario V90, albeit with much larger error. The asymmetry terms obtained from the three scenarios are listed in Tab. 1. This list includes 15 of the 17 observables that can be measured with a polarized beam and target. Missing are the tensor correlation coefficients $C_{yz,z}$ and $C_{xy,z}$, which would have required a dedicated run with guide fields as in scenario V45, but with longitudinal beam polarization.

Often angular distributions of the *same* asymmetry term (polarization times observable) are obtained from different scenarios. In addition, data have been collected during five runs, separated in time. Since the beam and target polarizations are not necessarily the same, these measurements may differ by an overall factor. We have checked that multiple measurements of the same asymmetry term (from different scenarios or from different runs), after normalization, are consistent with each other. We have also verified that the relative normalizations obtained from the analyzing powers are consistent with the (dependent) normalizations of the correlation coefficients. Multiple measurements of the same term are then averaged, resulting in an angular distribution for each asymmetry term.

## B. Beam and Target Polarization
### 1. General Remarks
In order to deduce the observables from the asymmetry terms, one must know the beam and target polarizations. This requires the determination of six numbers, namely $Q$, $P_\zeta$ and $P_{\zeta\zeta}$ at both beam energies.

To achieve this, we have used *two* sources of information, namely a global phase shift analysis of $p+p$ elastic scattering [24], and a published measurement of $p+d$ scattering with a 270 MeV polarized deuteron beam at RIKEN [3]. The normalization of all our data at both energies is based solely on these two data sets. This has been made possible by a series of auxiliary measurements as described in the following.

### 2. Vertical Beam Polarization at 135 and 200 MeV
At both energies a set of data is obtained with an unpolarized target, obtained by bleeding an $H_2$ - $D_2$ gas mixture into the target cell (Sec. III.4). The mixing ratio is adjusted to yield approximately the same number of $p+d$ and $p+p$ events. In addition to the normal sorting conditions for $p+d$ scattering events, a second set of conditions is used to select $p+p$ scattering events. Thus, the $p+d$ analyzing power $A_y^p$ and the $p+p$ analyzing power $A_y(pp)$ are measured simultaneously, with the *same* beam. The values of $A_y(pp)$ at the appropriate angles are obtained from the SAID phase shift solution SP03 [24]. Therefore, for this data sample, the beam polarization and consequently the $p+d$ analyzing power $A_y^p$ are known. This establishes a calibrated standard that can be used to deduce the beam polarization $Q$ from any data set that contains the asymmetry term $Q A_y^p$.

The statistical error that arises from normalizing the $p+p$ data to the phase shift solution is 0.9% at 135 MeV and 2.3% at 200 MeV.

### 3. Deuteron Target Polarization at 135 MeV
The vector and tensor analyzing powers for $p+d$ scattering have been measured recently at RIKEN [3] with 270 MeV deuterons, corresponding to a proton beam energy of 135 MeV. To obtain the RIKEN values at the angles measured in this experiment, we interpolate using a spline fit. The error of the interpolated values is taken as the average of the errors of the nearest-angle RIKEN points.

Scaling our asymmetry term $P_\zeta A_y^d$ to the RIKEN vector analyzing power $A_y^d$ yields the target vector polarization $P_\zeta$. After scaling, the two angular distributions are consistent. The statistical error of the normalization factor is 1.5%.

Scaling our asymmetry terms $P_{\zeta\zeta}A_\Delta$, $P_{\zeta\zeta}A_{zz}$ and $P_{\zeta\zeta}A_{xz}$ simultaneously to the corresponding RIKEN data yields the target tensor polarization $P_{\zeta\zeta}$. After scaling, the angular distributions for all three observables are consistent, with the exception of $A_\Delta$ at backward angles, which we thus exclude from the scaling procedure. The statistical error of the normalization factor is 1.9%.

### 4. Deuteron Target Polarization at 200 MeV

In order to transport the target polarization calibration from 135 MeV to 200 MeV, the Cooler is set up to accelerate the beam *during* an experimental cycle, a technique that has been described previously [25]. At the beginning of the cycle, unpolarized proton beam is injected at 135 MeV, and data are taken with a vector- and tensor-polarized target for about 100 s. The energy of the stored beam is then ramped to 200 MeV and data taking continues until the end of the cycle. This scenario is repeated for every cycle. It has been experimentally verified that the target polarization is constant during the ramp [25]. Since the target analyzing powers at 135 MeV are known [3], such a measurement calibrates the analyzing powers at 200 MeV.

For the calibration export only the forward angles, where the cross section is large, are used. The data at both energies are then scaled by the (common) target polarizations until they agree with the standard established at the lower energy. The statistical error of this normalization factor is 1.6% for the vector, and 2.4% for the tensor normalization. This results in calibrated deuteron analyzing powers at 200 MeV. The asymmetry terms of the main measurement at 200 MeV (at forward angles) are then scaled to the new standard. The error of this normalization is 1.2% for the vector, and 2.0% for the tensor normalization.

The combined normalization errors due to the target polarization at 200 MeV are then 2.0% for the vector, and 3.1% for the tensor part.

### 5. Longitudinal Beam Polarization

Data with longitudinal beam polarization have been obtained only at 135 MeV, and only with a vector-polarized target (scenario L90, Sec. IV.1.4).

The longitudinal beam polarization is determined from $p+p$ elastic scattering. Since the longitudinal analyzing power vanishes, spin correlation coefficients must be used and a polarized target is necessary. To this effect, the ABS is changed to produce a target of polarized H atoms.

The measured asymmetry terms $QC_{z,x}(pp)$ and $QC_{z,z}(pp)$ for $p+p$ scattering are then scaled simultaneously to the corresponding values of the SAID phase shift solution SP03 at the appropriate angles. The scaling error is 1.4%. This establishes the longitudinal beam polarization.

The $p+p$ data are bracketed in time by $p+d$ data runs immediately before and after. The measured asymmetry term $QA_y^p$ from the $p+d$ runs is the same within error, thus the beam polarizations $Q$ for the $p+p$ and the $p+d$ runs are also the same. The target vector polarization for scenario L90 is obtained as described in Sec. V.2.3, with a normalization error of 1.7%. The measured vector correlation coefficients can then be evaluated.

## C. Results

The normalization procedure described in the preceding section removes the polarizations from the asymmetry terms. At this stage, the terms containing more than one observable are reduced to single observables. The final results of this experiment are shown as solid symbols in Figs. 7 and 8. They are also available in numerical form from the authors upon request. The errors shown are statistical only. The corresponding normalization uncertainties are summarized in Tab. 2.

The open symbols in Figs. 7 and 8 mark previous polarization measurements in $p+d$ elastic scattering at or near the two energies of this experiment. A fairly large number of proton analyzing power data ($A_y^p$) have been measured; they include ref. [5] ($T_p = 135$ MeV, $31° < \theta < 170°$), ref. [26] ($T_p = 198$ MeV, $80° < \theta < 170°$), ref. [27] ($T_p = 120, 200$ MeV, $75° < \theta < 99°$), ref. [6] ($T_p = 135, 199$ MeV, $\theta = 94°$), and ref. [4] ($T_p = 190$ MeV, $30° < \theta < 115°$).

At $T_p = 135$ MeV, a comprehensive set of all four deuteron analyzing powers ($A_y^d$, $A_A$, $A_{zz}$ and $A_{xz}$), measured with a 270 MeV polarized deuteron beam, is reported in ref. [1] ($57° < \theta < 138°$) and refs. [2] and [3] ($10° < \theta < 66°$, $117° < \theta < 178°$). At $T_p = 200$ MeV, an older measurement of the deuteron analyzing powers ($35° < \theta < 135°$) exists [28]. Finally, the deuteron analyzing power ($A_y^d$) and the only previous spin correlation data ($C_{y,y}$) have been measured with an optically pumped target at the Indiana Cooler [7] ($T_p = 200$ MeV, $68° < \theta < 113°$).

Our data agree well with previous measurements, with the exception of the RIKEN measurement of $A_A$ at 135 MeV near $\theta \sim 155°$. Note that the normalization of the present data is independent of earlier measurements with the exception of the deuteron analyzing powers at 135 MeV [3] that were used to determine the target polarizations for the 135 MeV measurement.

## D. Cross Section

It is difficult to obtain a reliable figure for the absolute luminosity with an extended internal target and a stored beam, and thus a normalization for a cross section measurement. Nevertheless, it is still possible to extract a *relative* cross section, i.e., its angular dependence except for an unknown normalization factor. Agreement with existing data would then demonstrate that we understand our detector acceptance as a function of angle, and that any contributions from background are indeed negligible (Sec. IV.4).

To establish the detector efficiency, a Monte Carlo simulation is used, which contains a detailed account of all detector elements, including the silicon barrel, and describes the interaction of the reaction products with the detector setup to the best of our knowledge. Required input parameters include the detector positions, thicknesses and resolutions, the dimensions of the target cell and the target gas distribution. Also included is the loss of detected deuterons due to reactions in the forward detector (based on the parameterized total deuteron breakup cross section [29,30]). The simulation code produces output with the same format as that of the actual events recorded during data acquisition; therefore it can be analyzed with exactly the same software.

Elastic scattering events at random angles are processed by the Monte Carlo code and reconstructed with the same conditions as real events. The ratio between the number of reconstructed and generated events then constitutes the $\theta$-dependent detector efficiency $\varepsilon(\theta)$. The relative cross section is obtained by multiplying the measured yields by $\varepsilon(\theta)$. As a cross-check, the *relative* pp elastic scattering cross section can be determined from the data set obtained with the $H_2$ - $D_2$ gas mixture. It agrees well with the shape of the cross section predicted from the SAID phase shift solution SP03.

Our data at 135 MeV are shown as solid dots in Fig. 9. Two existing measurements by Ermisch et al. [31] (open circles) and Sekiguchi et al. [3] (stars) are in serious disagreement with each other in shape and magnitude. The shape of our cross section, in particular its forward/backward ratio, agrees well with the Ermisch data set, and is not compatible with the Sekiguchi measurement. We have thus normalized our cross section to the Ermisch data. In the past it has been argued that the minimum of the cross section is sensitive to three-nucleon forces [32]. For this reason, we also show in Fig. 9 a Faddeev calculation based on the CDBonn NN potential before (solid line) and after (dashed line) the inclusion of the Tucson-Melbourne three-nucleon force.

At 200 MeV (not shown) we have normalized our cross section to the data of Rohdjess et al. [33] at $\theta = 26^\circ$. The Rohdjess cross section is linked to $p+p$ scattering by the use of an HD gas target. With this normalization, our data are consistent at all angles with an older cross section measurement at 198 MeV [27].

We thus find that the shape of our cross section agrees well with existing data, without any correction for a background contribution. This supports our conclusion of Sec. IV.3 that background can be neglected.

## VI. Comparison with Theoretical Predictions
### A. Faddeev Calculations
The role of Faddeev calculations of observables involving three nucleons has recently been summarized by Glöckle in a comprehensive review [34]. Given a specific NN interaction as input, such calculations yield an exact solution of the three-body problem. Due to advances in computing power it is now possible to include a sufficient number of partial waves to extend these calculations up to ~200 MeV proton energy. Pion production is not included in these calculations.

The input NN interaction is represented by a modern NN potential whose parameters have been adjusted such that all empirical knowledge of the NN interaction is reproduced as well as possible. Such potentials are usually based on a parameterized one-boson exchange model with phenomenological parts added, and have been developed over the last 20 years. Following dramatic improvements in the past decade, modern potentials (including the so-called Bonn, Argonne and Nijmegen potentials I and II) yield a $\chi^2$ per datum of 1.0 to 1.4 for $p+p$ data up to 350 MeV, and 1.0 to 1.1 for $n+p$ data in the same energy range.

In this paper, we use Faddeev calculations that have been carried out by the Bochum-Cracow group [8], and are based on the following two NN potentials. The first, so-called CDBonn potential [35] has 45 free parameters, adheres most closely to a meson-exchange picture and is thus quite non-local. The second, the AV18 potential [36], is weakly non-local, has 40 free parameters and is more phenomenological than the CDBonn potential. Both potentials are charge dependent (i.e., not the same for $p+p$ and $n+p$), and the parameters of both have been adjusted by comparing to the Nijmegen NN phase shift analysis [37] at energies below 350 MeV.

The Faddeev calculations include the 3N partial wave states with total angular momenta of the two-nucleon subsystems up to $j_{max}=5$, resulting in up to 142 partial-wave states at each 3N system total angular momentum and parity. Convergence of observables for energies up to 200 MeV has been checked by comparing calculations with $j_{max}=5$ and $j_{max}=6$. Faddeev calculations ignore the Coulomb interaction. However, at our energies we expect Coulomb effects to be negligible, except perhaps at small angles. This is supported by experiment [38]. Thus, we assume that observables in $n+d$ and $p+d$ scattering are the same. On the other hand, Faddeev calculations are non-relativistic and use non-relativistic NN interactions. With increasing energy, relativistic effects become more important and may be responsible for some of the discrepancies between calculations and the data.

**B. Comparison of Two-Nucleon Force Predictions with the Data**
Our measured analyzing powers and spin correlation coefficients at 135 and 200 MeV are shown as solid circles in Figs. 7 and 8. Open symbols indicate the results of previous experiments (Sec. V.3). The solid and dashed lines show calculations with the CDBonn and the AV18 NN potential, respectively.

The ability of the calculations to account for the general behavior of all observables at both energies is quite impressive, especially since they were carried out before the data became available, and thus are true predictions. The difference between predictions of two potentials is generally small, as would be expected for NN potentials that have been adjusted to reproduce the NN database.

Discrepancies between the calculations and the data are mostly confined to backward angles but may be sizeable down to $\theta = 40^o$, especially in the tensor analyzing powers. Even though relatively small, these discrepancies are the focus of the present research, since they represent the physics that is missing in the 2N Faddeev calculations. The favored candidate for this physics is a three-nucleon force (3NF).

**C. Inclusion of a Three-Nucleon Force**

Most present day theoretical models of the 3NF are based on the exchange of two mesons with an intermediate nucleon excited state. There are two basic approaches. The first restricts the intermediate state to a $\Delta$ resonance and uses an additional, phenomenological, spin and isospin independent short-range part. An example is the Urbana IX force (UIX) [39]. The second approach is based on a parameterization of the π-N off-shell scattering amplitude and contains any intermediate state. A representative of the latter is the Tucson-Melbourne (TM) force [40]. Recently, The TM force has been criticized on the basis of chiral symmetry and a modified force (TM′) has been constructed that avoids these difficulties [41,42].

All three forces mentioned above have been adopted for insertion into Faddeev calculations [35], including angular momenta of the 3N system up 13/2 [8]. All theoretical 3NFs contain adjustable parameters that are determined experimentally. In particular, the overall strength of the 3NF potential is adjusted by varying the cut-off parameter $\Lambda$ of the π-N form factor until the $^3$H binding energy is reproduced. The adjusted cut-off parameter depends on the NN potential used [43].

**D. Comparison of 3NF Predictions with the Data**

The differences between our measurements and the Faddeev calculation with the CDBonn potential are plotted in Figs. 10 and 11, i.e., the calculation is the zero line. The effect of including the old (TM) or the new (TM′) Tucson-Melbourne 3NFs is shown by the solid lines and the dashed lines respectively. A comparison of these curves with the data is justified if calculations with different NN potentials agree with each other. To illustrate this, the difference between calculations with the AV18 and the CDBonn potentials, both without a 3NF, is shown as a dotted line. This difference is indeed generally small, but there are many cases where the variation between the two potentials competes in size with the 3NF effects.

As can be seen from Figs. 10 and 11, the two 3NFs agree with each other for some observables and in some angular regions (e.g., in $A_y^d$), but in numerous cases the predictions with the TM and the TM′ 3NF are quite different. Both sometimes improve the agreement with the data (e.g., in $A_y^d$), but equally often this is not the case. Thus, neither 3NF is a successful representation of the discrepancies between the $p+d$ spin observables and Faddeev calculations without a 3NF.

In Fig. 12 we investigate the systematics of the performance of various 3NFs and underlying NN potentials. Each panel shows the measured observables versus the scattering angle, thus each pixel corresponds to one of our 868 data points. A pixel is colored black if the inclusion of a 3NF *improves* the agreement with the data and gray if it doesn't. The top four panels are for 135 MeV, the lower four for 200 MeV. The left column is with the CDBonn NN potential (TM or TM′), the right with the AV18 (TM or UIX). It is interesting to note that there are no systematic differences between different regions in scattering angle, different 2N potentials, or different 3NFs.

In summary, there is no indication that any of the 3NFs studied here consistently alleviates the discrepancies between the data and 2N Faddeev calculations, and thus represents the physics that is responsible for these discrepancies.

## VII. Conclusions

We have measured all analyzing powers, and all but two spin correlation coefficients for *p+d* elastic scattering at 135 and 200 MeV. The experiment was motivated by the availability of computationally exact Faddeev calculations of these observables. These calculations are based on a given, phenomenological 2N potential.

The Faddeev calculations shown in this paper were carried out prior to this experiment. We find that the 2N calculations predict the general features of all observables impressively well. In other words, the absolute differences between data and the two-nucleon force calculations are relatively small, mostly confined to backward angles but in some cases sizeable down to $\theta = 40^\circ$. Statistically, the discrepancies are relatively large owing to the high precision of the data. If the 2N input to the calculation is sufficiently well defined, such that it uniquely describes how nature would behave if there were only 2N forces, the differences between these calculations and the data are a manifestation of additional physics. Our measurement then would provide a testing ground for the spin dependence of this missing physics.

Many believe that the prime candidate for the missing physics is a three-nucleon force. It is possible to include theoretical models of three-nucleon potentials in the Faddeev calculations. We have investigated the ability of three different three-nucleon forces to account for the discrepancies between data and 2N calculations. We find that for some observables at some angles the inclusion of a 3NF improves the agreement with the data, but often the agreement also gets worse. When there is an improvement, it does not depend systematically on the scattering angle, or the energy, or the choice of a particular 3NF. We thus conclude that existing 3NFs are not successful in explaining the discrepancy between the spin observables presented here and the corresponding 2N calculations. Thus, recent claims that local improvements of the calculation resulting from inclusion of a 3NF constitute evidence for such a 3NF must be met with caution. For example, in Ref. [7], that claim is based on a (fortuitous) choice of a single observable ($C_{y,y}$) in a limited angular range (the *data* of ref. [7] are in agreement with the present measurement, but the conclusion is not).

We have also resolved a serious discrepancy between two recent measurements of the differential cross section at 135 MeV (Sec. V.4).

## Acknowledgements

We would like to thank Terry Sloan, Gary East, and the members of the Operations Group, Brian Allen, Pete Goodwin, Glen Hendershot, Mark Luxnat, Tom Meaden, and John Ostler for their hard work at odd hours that made this experiment possible. We also appreciate the help of Joanna Kuroś-Żołnierczuk with the comparisons to theory, and we would like to thank Tom Finnessy for his help in the production of the transition units. We are grateful to the authors of Ref. [8] for making available to us Faddeev calculations obtained with their model. The Faddeev calculations were carried out on the Cray T90, SV1 and T3E of the NIC in Jülich, Germany. We are grateful to Walter Glöckle for a careful reading of the manuscript. This work has been carried out under NSF grant PHY-0100348, and DOE grant FG02-88ER40438.

[†] deceased

| term | scenario | | | term | Scenario | | | term | scenario | | |
|---|---|---|---|---|---|---|---|---|---|---|---|
| | V90 | V45 | L90 | | V90 | V45 | L90 | | V90 | V45 | L90 |
| $QA_y^p$ | v | v | v | $QP_\zeta (C_{x,x}+C_{y,y})$ | v | | v | $QP_{\zeta\zeta} C_{zz,y}$ | v | v | |
| $P_\zeta A_y^d$ | v | | v | $QP_\zeta (C_{x,x}-C_{y,y})$ | v | | v | $QP_{\zeta\zeta} C_{\Delta,y}$ | v | | |
| $P_{\zeta\zeta} A_\Delta$ | v | v | | $QP_\zeta C_{z,x}$ | v | | v | $QP_{\zeta\zeta} C_{xz,y}$ | | v | |
| $P_{\zeta\zeta} A_{zz}$ | v | v | | $QP_\zeta C_{x,z}$ | | | v | $QP_{\zeta\zeta} C_{yz,x}$ | | v | |
| $P_{\zeta\zeta} A_{xz}$ | | v | | $QP_\zeta C_{z,z}$ | | | v | $QP_{\zeta\zeta} (C_{xy,x}-\frac{1}{2}C_{\Delta,y})$ | v | v | |
| | | | | | | | | $QP_{\zeta\zeta}(C_{xy,x}+\frac{1}{2}C_{\Delta,y})$ | v | v | |

**Table 1: List of asymmetry terms obtained under the different running conditions (scenarios). For details see Sects.IV.1 and V.2.**

| Energy (MeV) | 135 | 200 |
|---|---|---|
| $A_y^p$ | 0.9 | 2.3 |
| $A_y^d$ | 1.5 | 2.0 |
| $A_{xz}, A_{yy}, A_{zz}$ | 1.9 | 3.1 |
| Vector correlation coefficients | 1.7 | 3.0 |
| Tensor correlation coefficients | 2.1 | 3.9 |
| $C_{x,z}, C_{z,z}$ | 4.6 | - |

**Table 2: Overall normalization errors in % for the different observables at 135 and 200 MeV (for more detail, see SectV.2)**

# Appendix

In Sec.II.2 we have discussed the derivation of the spin-dependent cross section for scenario V90. Here we give the corresponding expressions that apply in case of the other two scenarios used (Sec. IV.1).

For scenario L90, the beam polarization $\hat{Q}$ is longitudinal ($\beta_Q = 0$). For sideways spin alignment axis $\hat{S}$ ($\beta_P = \pi/2$, $\Phi_P = 0$), we then obtain

$$\sigma/\sigma_0 = 1 - \tfrac{3}{2} P_\varsigma A_y^d \sin\varphi - \tfrac{1}{4} P_{\varsigma\varsigma}[A_{zz} - A_\Delta \cos 2\varphi]$$
$$+ \tfrac{3}{2} P_\varsigma Q C_{x,z} \cos\varphi - \tfrac{1}{2} P_{\varsigma\varsigma} Q C_{xy,z} \sin 2\varphi , \qquad (A.1)$$

with a vertical spin alignment axis ($\beta_P = \pi/2$, $\Phi_P = \pi/2$),

$$\sigma/\sigma_0 = 1 + \tfrac{3}{2} P_\varsigma A_y^d \cos\varphi - \tfrac{1}{4} P_{\varsigma\varsigma}[A_{zz} + A_\Delta \cos 2\varphi]$$
$$+ \tfrac{3}{2} P_\varsigma Q C_{x,z} \sin\varphi + \tfrac{1}{2} P_{\varsigma\varsigma} Q C_{xy,z} \sin 2\varphi , \qquad (A.2)$$

and with a longitudinal spin alignment axis ($\beta_P = 0$),

$$\sigma/\sigma_0 = 1 + \tfrac{1}{2} P_{\varsigma\varsigma} A_{zz} + \tfrac{3}{2} P_\varsigma Q C_{z,z} \qquad (A.3)$$

For scenario V45, the beam polarization was vertical, and the longitudinal and one of the transverse guide fields was energized simultaneously. The sub-cycle covered all eight possible orientations of the spin alignment axis. When combining the *transverse* with the *longitudinal* field, the following four spin alignment axis directions result

$$(\beta_P, \Phi_P) = \begin{bmatrix} (3\pi/4, 0) & (\pi/4, 0) \\ (3\pi/4, \pi) & (\pi/4, \pi) \end{bmatrix} \qquad (A.4)$$

The corresponding four cross sections are the same except for the signs of the terms. The signs in the following equation are shown as matrices that correspond to the directions of Eq. A.4.

$$\sigma/\sigma_0 = 1 \begin{bmatrix} + & + \\ + & + \end{bmatrix} Q A_y^p \cos\varphi \begin{bmatrix} - & - \\ + & + \end{bmatrix} \tfrac{3\sqrt{2}}{4} P_\varsigma A_y^d \sin\varphi \begin{bmatrix} - & + \\ + & - \end{bmatrix} \tfrac{1}{2} P_{\varsigma\varsigma} A_{xz} \cos\varphi$$
$$\begin{bmatrix} + & + \\ + & + \end{bmatrix} \tfrac{1}{8} P_{\varsigma\varsigma} A_\Delta \cos 2\varphi \begin{bmatrix} + & + \\ + & + \end{bmatrix} \tfrac{1}{8} P_{\varsigma\varsigma} A_{zz} \begin{bmatrix} + & + \\ - & - \end{bmatrix} \tfrac{3\sqrt{2}}{8} P_\varsigma Q \{C_{x,x} - C_{y,y}\} \cos 2\varphi$$

$$\begin{bmatrix} - & + \\ - & + \end{bmatrix} \tfrac{3\sqrt{2}}{4} P_{\varsigma} Q C_{z,x} \sin\varphi \begin{bmatrix} - & - \\ - & - \end{bmatrix} \tfrac{1}{8} P_{\varsigma\varsigma} Q\{C_{xy,x} - \tfrac{1}{2} C_{\Delta,y} - C_{zz,y}\} \cos\varphi \qquad (A.5)$$

$$\begin{bmatrix} - & + \\ + & - \end{bmatrix} \tfrac{1}{4} P_{\varsigma\varsigma} Q\{C_{xz,y} + C_{yz,x}\} \cos 2\varphi \begin{bmatrix} + & + \\ + & + \end{bmatrix} \tfrac{1}{8} P_{\varsigma\varsigma} Q\{C_{xy,x} + \tfrac{1}{2} C_{\Delta,y}\} \cos 3\varphi$$

$$\begin{bmatrix} - & + \\ + & - \end{bmatrix} \tfrac{1}{4} P_{\varsigma\varsigma} Q\{C_{xz,y} - C_{yz,x}\}$$

When combining the *vertical* with the *longitudinal* guide field, the following four spin alignment axis directions result

$$(\beta_P, \Phi_P) = \begin{bmatrix} (3\pi/4, \pi/2) & (\pi/4, \pi/2) \\ (3\pi/4, 3\pi/2) & (\pi/4, 3\pi/2) \end{bmatrix} . \qquad (A.6)$$

and the corresponding four cross sections are

$$\sigma/\sigma_0 = 1 \begin{bmatrix} + & + \\ + & + \end{bmatrix} Q A_y^p \cos\varphi \begin{bmatrix} + & + \\ - & - \end{bmatrix} \tfrac{3\sqrt{2}}{4} P_{\varsigma} A_y^d \cos\varphi \begin{bmatrix} - & + \\ + & - \end{bmatrix} \tfrac{1}{2} P_{\varsigma\varsigma} A_{xz} \sin\varphi$$

$$\begin{bmatrix} - & - \\ - & - \end{bmatrix} \tfrac{1}{8} P_{\varsigma\varsigma} A_{\Delta} \cos 2\varphi \begin{bmatrix} + & + \\ + & + \end{bmatrix} \tfrac{1}{8} P_{\varsigma\varsigma} A_{zz} \begin{bmatrix} + & + \\ - & - \end{bmatrix} \tfrac{3\sqrt{2}}{8} P_{\varsigma} Q(\{C_{x,x} + C_{y,y}\} - \{C_{x,x} - C_{y,y}\} \cos 2\varphi)$$

$$\begin{bmatrix} - & + \\ - & + \end{bmatrix} \tfrac{3\sqrt{2}}{4} P_{\varsigma} Q C_{z,x} \sin\varphi \begin{bmatrix} + & + \\ + & + \end{bmatrix} \tfrac{1}{8} P_{\varsigma\varsigma} Q\{C_{xy,x} - \tfrac{1}{2} C_{\Delta,y} + C_{zz,y}\} \cos\varphi \qquad (A.7)$$

$$\begin{bmatrix} - & + \\ + & - \end{bmatrix} \tfrac{1}{4} P_{\varsigma\varsigma} Q\{C_{xz,y} + C_{yz,x}\} \sin 2\varphi \begin{bmatrix} - & - \\ - & - \end{bmatrix} \tfrac{1}{8} P_{\varsigma\varsigma} Q\{C_{xy,x} + \tfrac{1}{2} C_{\Delta,y}\} \cos 3\varphi .$$

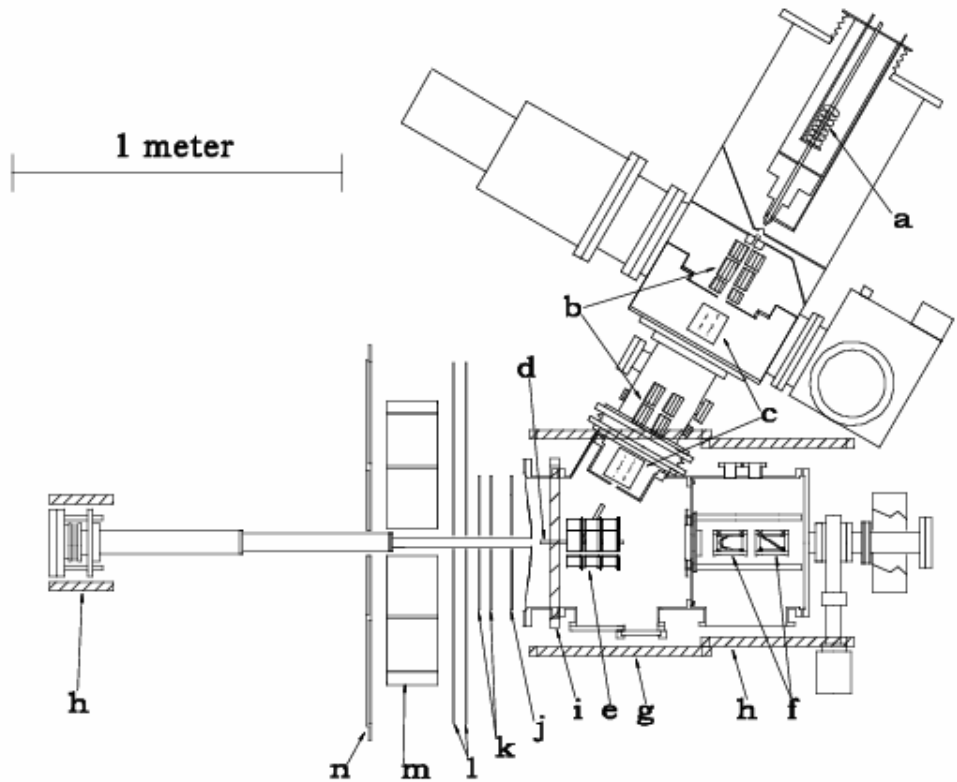

**FIG. 1.** Top view of the target and detector setup. The stored beam travels from right to left. Shown are the atomic beam source and the target cell (a-d), the detector system (e, j-m), and the guide field (i,g) and compensating (h) coils. An additional 6.4 mm thick scintillator detector (n) is not used in this experiment. Also shown are two beam position monitors (f).

**FIG. 2.** Array of 18 micro-strip recoil detectors (Silicon Barrel). Also shown is the

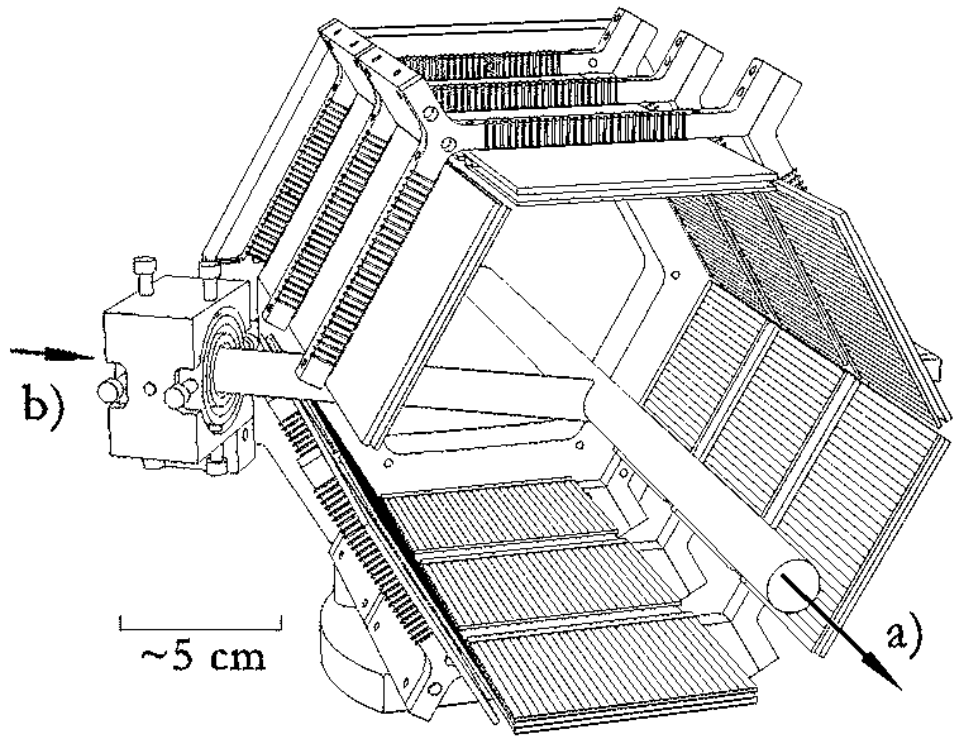

thin-walled target cell. The direction of the stored beam (a), and the direction of the polarized atomic beam (b) are indicated.

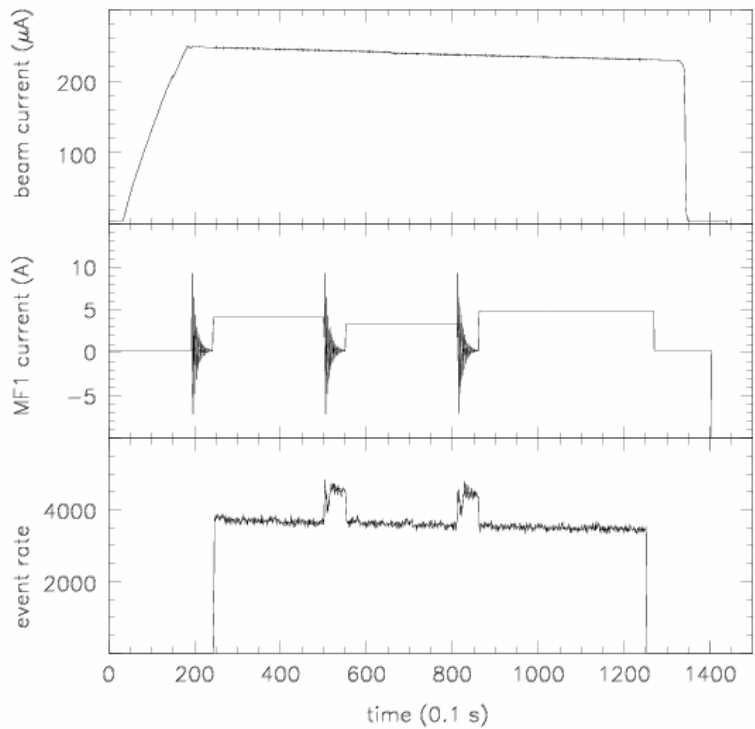

**FIG. 3.** Stored beam current, the current in the MF1 offset coil, and the event rate during data taking during a scenario-V90 cycle. The cycle length is 140 s. The increases in event rate are due to the thicker target during the degaussing of the transition units.

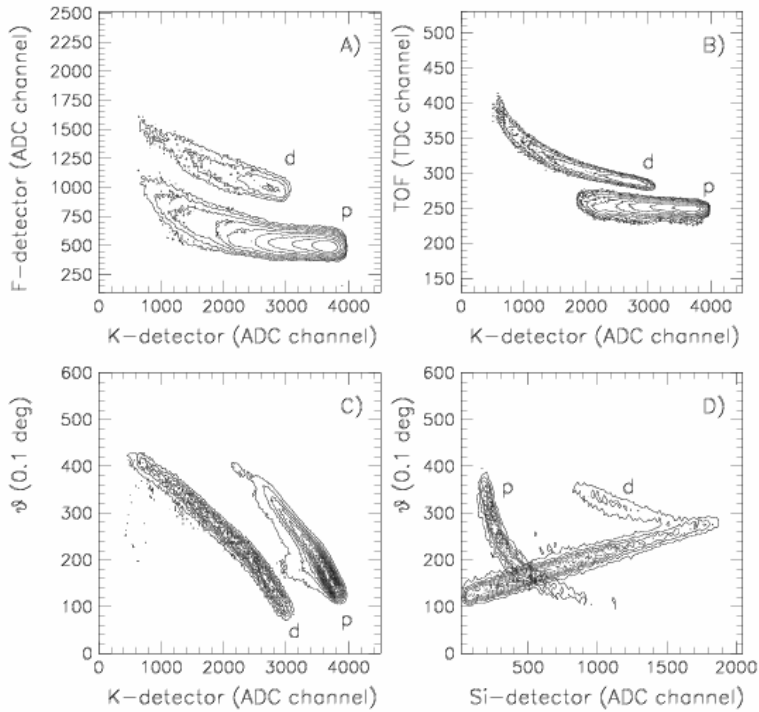

**FIG. 4. Identification of elastic scattering events at 135 MeV. Since the cross section for the two cases is very different, the contour values have been adjusted separately. Panels A – C show the energy in the F detector, the time-of-flight between the F and the K detector and the angle of the forward prong versus the energy deposited in the K detector (in arbitrary units). The forward angle versus the energy of the recoil is shown in panel D. The loci corresponding to a forward-going proton or deuteron are labeled accordingly.**

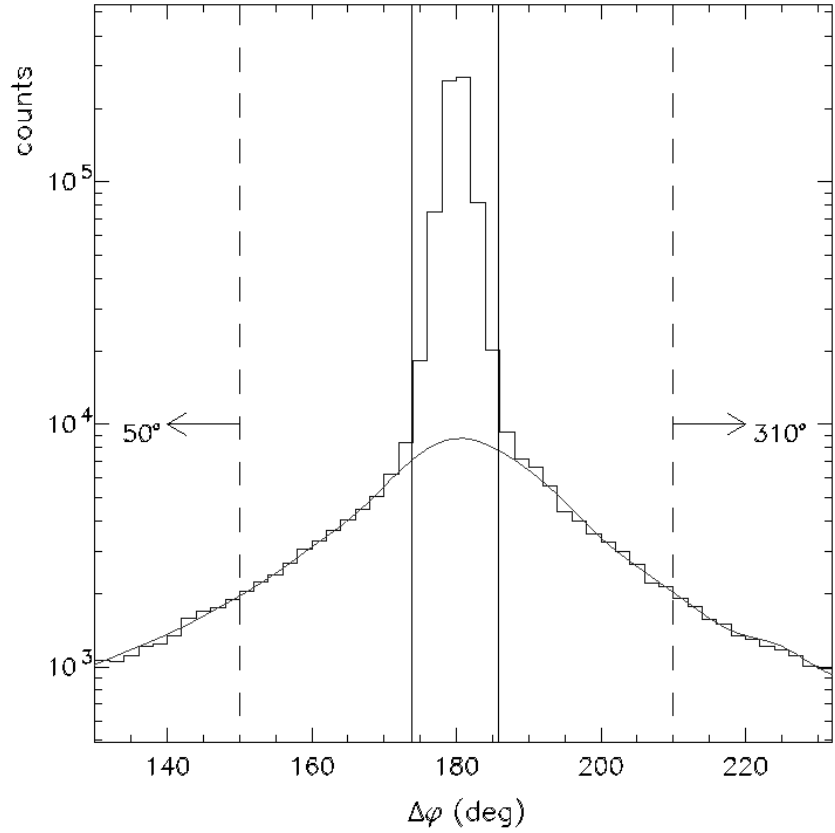

**FIG. 5.** Distribution of Δφ, the difference between the azimuth of the forward and the recoil particle. The peak at $180^\circ$ is due to (coplanar) elastic scattering. Gates used for real event and background identification are indicated by the solid and dashed lines respectively. The effect of the background (solid line) on the data is discussed in Sec. IV.3.

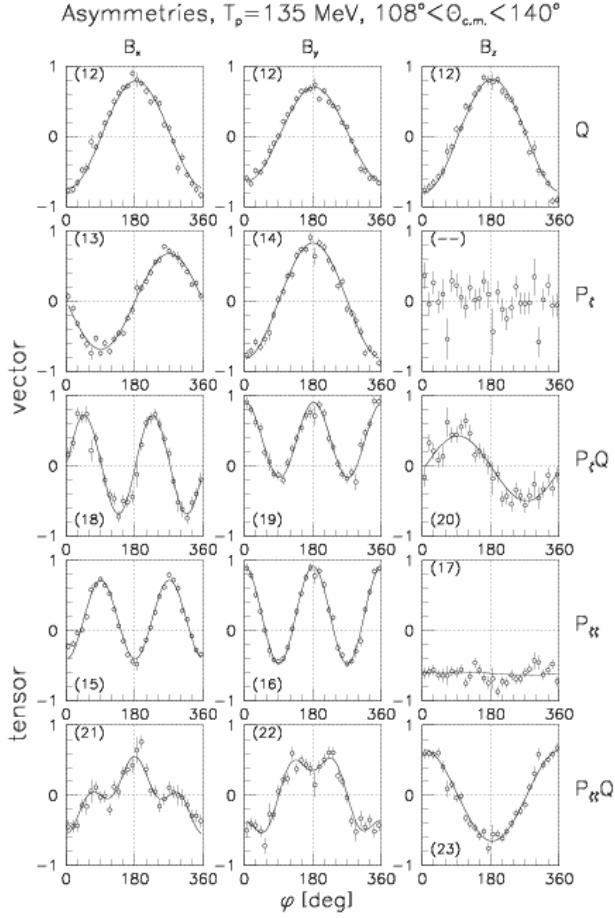

**FIG. 6.** The asymmetries R for scenario V90 versus the azimuth $\varphi$. The three columns correspond to the orientations x, y and z of the target guide field. The five rows are for the target asymmetry, the vector target and vector correlation asymmetries, and the tensor target and tensor correlation asymmetries. The numbers in brackets refer to the corresponding equations in Sec. V.1.3; the fit based on these expressions is shown as a line. The values of R are scaled to fill the graphs. Scale factors range from 3.8 to 13.0. For this figure, polar angles from $108°$ to $140°$ have been integrated. The asymmetry in the unnumbered panel is expected to vanish by parity conservation.

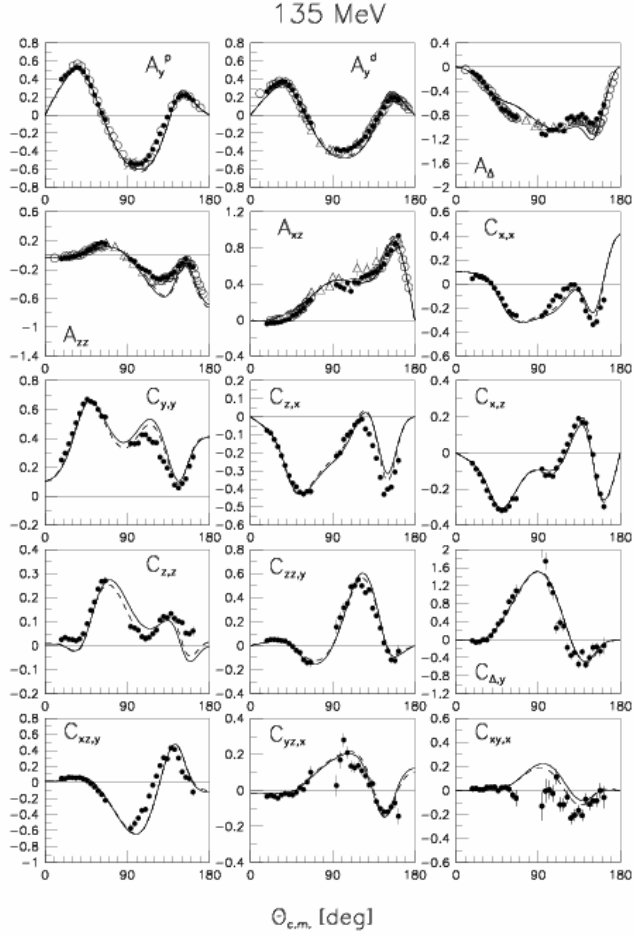

**FIG. 7.** Spin observables for *p+d* elastic scattering at Tp = 135 MeV. The solid dots represent the results of this experiment. Statistical errors are shown; the overall normalization errors are listed in Tab. 2. The open symbols show previous measurements (Sec. V.3). The solid and dashed curves are two-nucleon force Faddeev calculations based on the CD-Bonn and the AV18 NN potential, respectively (Sec. VI.2).

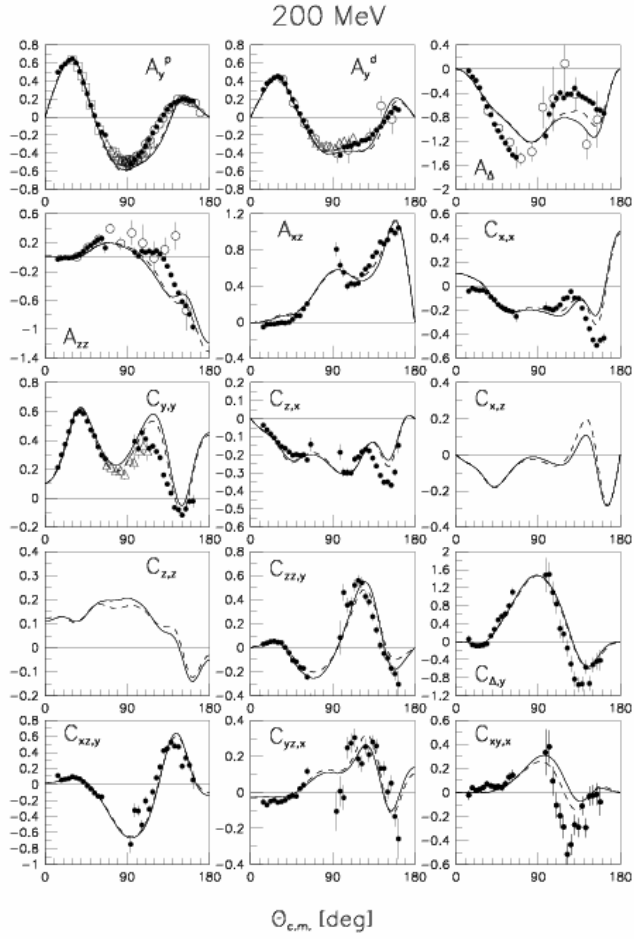

**FIG. 8.** Spin observables for *p+d* elastic scattering at $T_p$ = 200 MeV. Otherwise, the caption of Fig. 7 applies.

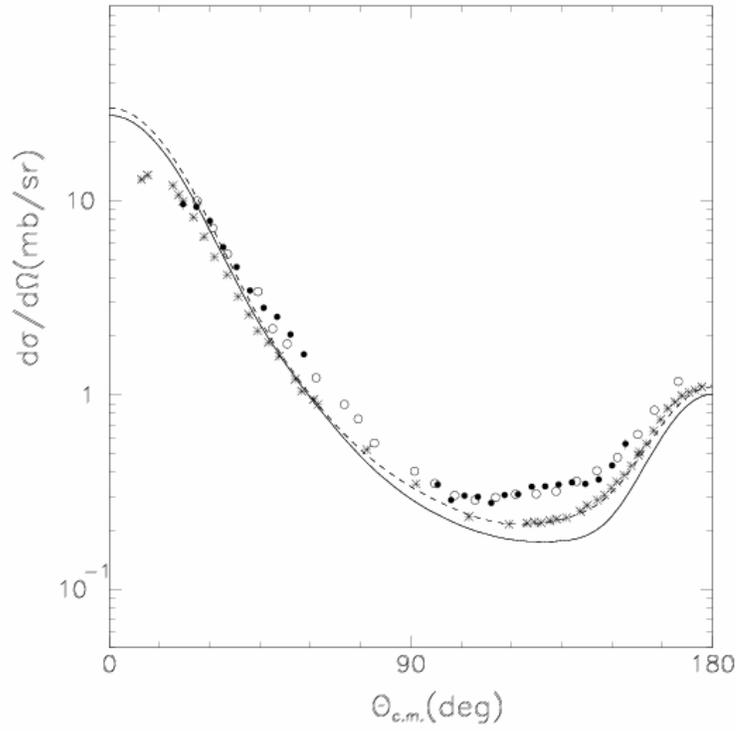

FIG. 9. Differential cross section for pd elastic scattering at 135 MeV. The relative cross sections of this experiment (solid dots) are normalized to the data by Ermisch et al. [31] (open circles). Also shown is another recent measurement [3] (stars), which is in conflict with the other data. The solid line represents a Faddeev calculation based on the CDBonn NN potential; when the TM three-nucleon force is included, the dashed line results.

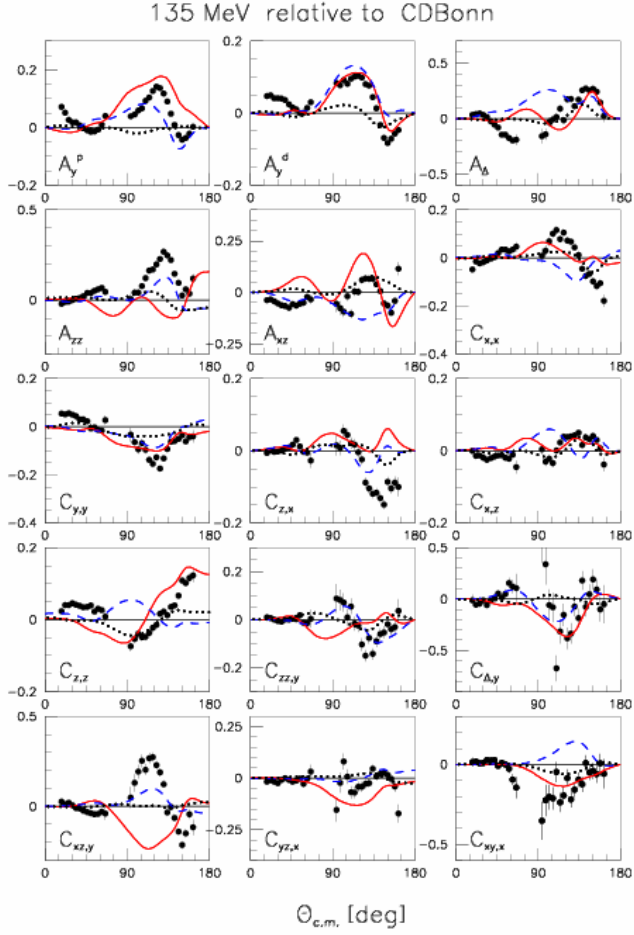

FIG. 10. (Color online) Difference between the present data at 135 MeV and the Faddeev calculation with the CDBonn potential. The effect of including the old or the new Tucson-Melbourne 3NFs is shown by the solid lines (TM) and the dashed lines (TM′). The dotted lines show the difference between calculations with the AV18 and the CDBonn potentials, both without a 3NF.

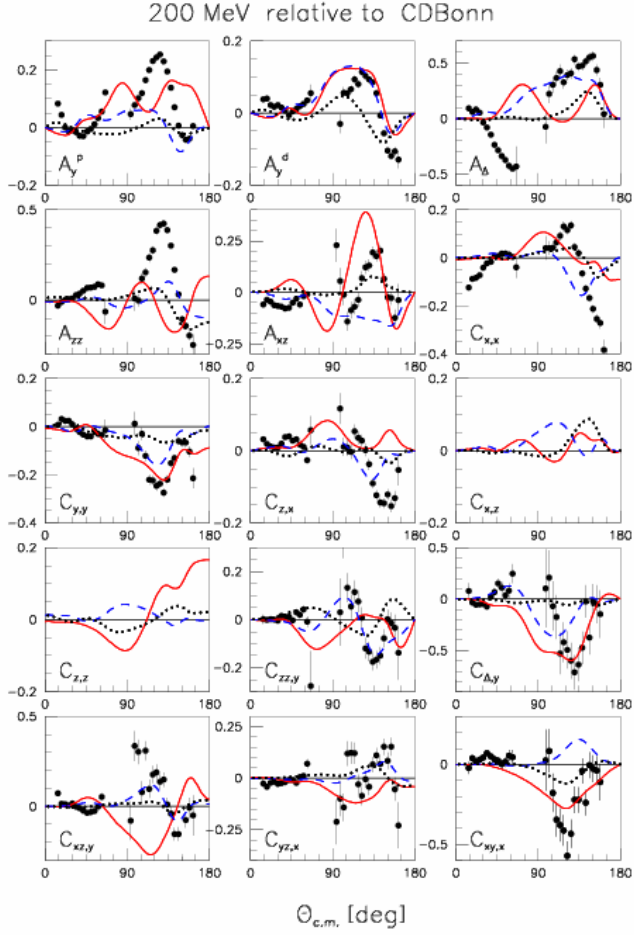

**FIG. 11. (Color online) Difference between the present data at 200 MeV and the Faddeev calculation with the CDBonn potential. Otherwise, the caption of Fig. 10 applies.**

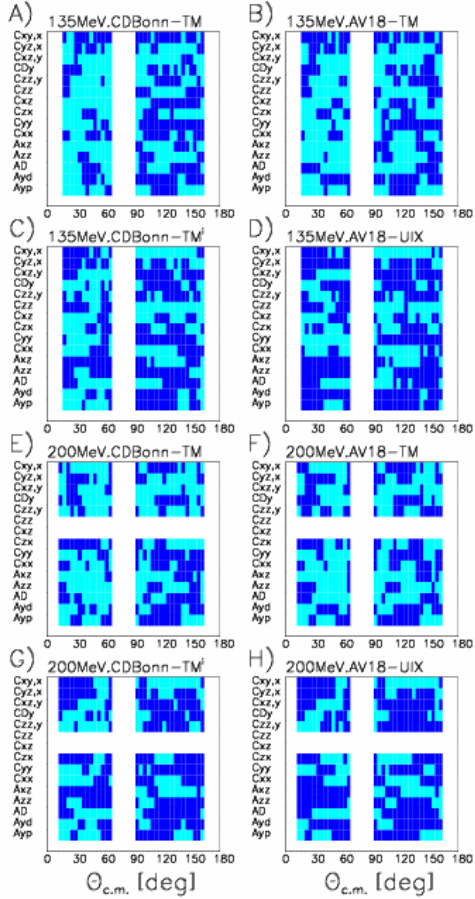

FIG. 12. (Color online) Systematics of including various three-nucleon forces. Each panel shows the measured observable versus the scattering angle, thus each pixel corresponds to one of our 868 data points. A pixel is colored black if the inclusion of a 3NF improves the agreement with the data. The upper four panels are for 135 MeV, the lower four for 200 MeV. The left and right columns are for the CD-Bonn and the AV18 2N force, respectively. The effect of three different 3N forces is shown.